\documentclass[%
 reprint,
 amsmath,amssymb,
 aps,
]{revtex4-2}

\usepackage{graphicx}
\usepackage{dcolumn}
\usepackage{bm}
\usepackage{epstopdf}
\usepackage{hyperref}
\makeatletter

\newcommand{\ZZ}{\mathbb{Z}}
\newcommand{\xiu}{\xi_u}
\newcommand{\Lmin}{L_{\rm min}}
\newcommand{\zc}{z_{\rm c}}
\newcommand{\scrO}{\mathcal{O}}
\newcommand{\PP}{\mathbb{P}}
\newcommand{\var}{{\rm Var}}
\newcommand{\EE}{\mathbb{E}}
\newcommand{\scrN}{\mathcal{N}}
\newcommand{\scrD}{\mathcal{D}}
\newcommand{\scrR}{\mathcal{R}}
\usepackage{caption}
\begin{document}

\preprint{APS/123-QED}

\title{Self-avoiding trails in two and three dimensions}

\author{Xiaodi Su$^{1}$}
\author{Zongzheng Zhou$^{2}$}%
\author{Qianqian Wu$^{1,*}$}
\email{qian-qian.wu@hfut.edu.cn}

\affiliation{$^{1}$School of Mathematics, Hefei University of Technology, Hefei, Anhui 230009, China}
\affiliation{$^{2}$School of Mathematics, Monash University, Clayton, Victoria 3800, Australia}


\begin{abstract}
The self-avoiding trail is an important variant of the self-avoiding walk. In this work, we employ an irreversible Markov chain Monte Carlo algorithm, together with the reversible Berretti-Sokal algorithm for comparison, to simulate self-avoiding trails on the square and simple cubic lattices with periodic boundary conditions. Based on finite-size analyses of the unwrapped end-to-end distance and the Binder ratio, we accurately estimate the critical points on the simple cubic and square lattices to be 0.206\,376\,9(2) and 0.367\,561\,1(1), respectively, improving the precision of previous best estimates by factors of 500 and 70. At the estimated critical point of the simple cubic lattice, we numerically demonstrate that both the critical scaling behaviors of various quantities and the length distributions of self-avoiding trails and walks are consistent with each other. Our accurate numerical results are attributed to the efficiency of the irreversible algorithm, whose advantage over the reversible algorithm is even more pronounced for the self-avoiding trail model than for the self-avoiding walk model.
\begin{description}
\item[Keywords]
Irreversible algorithm, self-avoiding trails, critical points
\end{description}
\end{abstract}

\maketitle


\section{\label{sec:level1}INTRODUCTION}
Lattice path models, originally introduced to describe polymer chains in solution~\cite{Flory1949, Orr1947}, are fundamental systems in statistical mechanics for studying phase transitions. Among them, two of the most studied models are self-avoiding walks (SAWs)~\cite{madras2013self} and self-avoiding trails (SATs)~\cite{malakis1975self,shapir1984walks}. The former prohibits repeated visits to the same lattice site, whereas the latter prohibits multiple traversals along the same lattice edge. 
On the $d$-dimensional lattice ($\ZZ^d$), the generating functions for SAWs and SATs, starting at the origin, can be written as~\cite{madras2013self}
\begin{equation}
\label{Eq:chi-gen}
\chi(z)=\sum_{\mathcal{N}=0}^{\infty}c_\mathcal{N}z^\mathcal{N}=\sum_{\omega}z^{|\omega|},
\end{equation}
where $z>0$ is the fugacity, $c_\mathcal{N}$ denotes the number of SAWs or SATs of length $\mathcal{N}$, and the second summation is taken over all admissible walks or trails $\omega$. Here, $|\omega|$ denotes the length of $\omega$, i.e., the number of edges on the walk or trail. This admits a probabilistic interpretation: each admissible walk/trail $\omega$ is assigned with the weight $z^{|\omega|}$. For $d\geq 2$, it is expected that
\begin{equation}
    c_\mathcal{N}\sim \mu^{\mathcal{N}}\mathcal{N}^{\gamma-1}\;,
\end{equation}
where $\mu$ is called the connective constant, and $\gamma$ is a critical exponents~\cite{madras2013self}. It is also expected that $\chi(z)$ has a positive radius of convergence $z_c$ which relates to $\mu$ as $z_c = 1/\mu$. Near the critical point $z_c$, a phase transition happens. Define the mean end-to-end distance as $\xi(z)=\sqrt{\sum_{\omega}z^{|\omega|}\|x_{\omega}\|^2/\sum_{\omega}z^{|\omega|}}$. For $z<z_c$, the mean walk/trail length is finite, whereas it is infinite when $z \geq z_c$. As $z$ approaches $z_c$ from below, both $\chi(z)$ and $\xi(z)$ diverge as power law,
\begin{equation}
    \chi(z)\sim (z_c-z)^{-\gamma}, \xi(z)\sim (z_c-z)^{-\nu},
\end{equation}
where $\gamma, \nu$ are critical exponents. 




SAWs have been extensively studied over the past several decades. On the honeycomb lattice, the critical point of the SAW was firstly conjectured as $z_c =1/\sqrt{2+\sqrt{2}}$ by Nienhuis in 1982~\cite{nienhuis1982exact} and then after 30 years was rigorously proved by Duminil-Copin and Smirnov~\cite{duminil2012connective}. High-precision estimates of $z_c$ are available for several two-dimensional and higher-dimensional lattices, and an introductory review can be found in Ref.~\cite{fang2021logarithmic}. For example, on the simple cubic lattice, it was obtained that $z_c =0.213 491 0(3)$. The critical behavior of the SAW is characterized by universal critical exponents. In two dimensions, it was conjectured that $\gamma=43/32$ and $\nu=3/4$. In three dimensions, numerical studies have determined these exponents to a surprisingly high precision, yielding $\gamma=1.156\,953\,00(95)$ and $\nu=0.587\,597\,00(40)$~\cite{jensen2004self}. The upper critical dimension of SAW is $d=4$, and for $d\geq 4$ critical exponents are expected to take mean-field values $\gamma = 1$ and $\nu = 1/2$~\cite{hara1992self}.

The SAT was first proposed by Malakis in 1975~\cite{malakis1975self,malakis1976trail}. Using exact enumeration, Malakis found that on the square lattice several critical exponents for the SAT are consistent with those of the SAW. In 1984, Zhou and Li~\cite{ZhouLi1984} suggested, using series analysis and real-space renormalization group methods, that the SAT might not belong to the same universality class as the SAW. However, in 1985, Guttmann~\cite{guttman1985lattice} showed that the SAW and SAT belong to the same universality class on the two-dimensional honeycomb lattice and on the three-dimensional Lave's lattice. In particular, on the honeycomb lattice, the connective constant of SAT was obtained as $\mu = \sqrt{2+\sqrt{2}}$, which coincides with the connective constant of SAW on the honeycomb lattice. In the same year, using series analysis, Guttmann estimated that the connective constants on the square, triangular and simple cubic lattices were $2.721\,5(20)$, $4.524(4)$ and $4.843(3)$~\cite{guttmann1985lattice}. Guttmann and Osborn~\cite{guttmann1988monte} subsequently used the Berretti-Sokal algorithm to obtain more accurate connective constants with $\mu=2.720\,59(42)$ for the square lattice and $\mu=4.525\,26(15)$ for the triangular lattice. They also argued that this Monte Carlo method is more efficient than series analysis for studying trail models. In 1989, Meirovitch and Lim~\cite{meirovitch1989computer} studied the SAT on the square lattice by the scanning simulation method, and estimated $\mu = 2.720\,58(20)$ and $\gamma = 1.350(12)$. Finally, Conway and Guttmann~\cite{conway1993enumeration} enumerated the SATs using a transfer matrix technique, and obtained the critical point $z_c=0.367\,562(7)$ on the square lattice.

Compared with SAW, the estimates of SAT critical points on the square and simple-cubic lattices are less accurate, especially for the simple-cubic lattice case. Therefore, in this paper we perform a systematic study to the SAT on the square and simple cubic lattices, aiming to improve the precision of estimates of the critical points. We employ an efficient irreversible Monte Carlo algorithm to simulate SATs on the square and simple-cubic lattices with periodic boundary conditions (PBC). For the SAW model, this algorithm has been shown to be more efficient than traditional reversible algorithms like the Berretti-Sokal (BS) algorithm and Metropolis-Hastings algorithm. Remarkably, for the SAT model, we find that the efficiency advantage of the irreversible algorithm is even more pronounced than in the SAW case. By simulating systems with linear sizes up to $L=128$ on simple cubic lattices, we obtain a precise estimate of $z_c = 0.206\,376\,9(2)$, from a finite-size scaling analysis of the critical behavior of a Binder ratio and unwrapped end-to-end distance. This result improves the previously best estimate by about 500 times. We also estimate the critical exponents to be $\nu=0.589\,3(7)$ and $\gamma=1.158(2)$. In two dimensions, by simulating systems with sizes up to $L=512$, the critical point is estimated as $z_c=0.367\,561\,1(1)$, representing an improvement of nearly 70 times over the best result in Ref.~\cite{conway1993enumeration}.

The remainder of this article is organized as follows. In Section~\ref{sec:model}, we introduce a reversible and an irreversible Markov chain Monte Carlo algorithms; both are used to simulate SAT and their efficiency is discussed in Section~\ref{sec:re}. The sampled quantities and their expected finite-size scaling are summarized in Section~\ref{Sec:Observables}. In Section~\ref{sec:re}, we provide a detail analysis for estimating critical points of SAT on square and simple-cubic lattices, as well as the critical exponents on the simple-cubic lattice. This paper is concluded in Section~\ref{Sec:Conclusion}.

\section{Algorithms}
\label{sec:model}

In this work, we simulate SATs on square and simple cubic lattices with periodic boundary conditions, and all trails are fixed to start from the origin of the lattice. We employ the BS algorithm and an irreversible algorithm. These two algorithms are used to study self-avoiding walks in Ref.~\cite{hu2017irreversible}.


We first introduce the BS algorithm~\cite{berretti1985new}. Consider a self-avoiding trail with one endpoint (called the tail) fixed at the origin of the lattice, and let $\scrN$ denote the current length of the trail. The other endpoint is called the head. The BS algorithm updates the SAT according to the following steps.
\begin{enumerate}
    \item Choose uniformly at random one of the two actions: ``Add" or ``Delete".
    \item If the ``Add" action is selected, then uniformly at random choose one of the $2d-1$ neighboring sites of the head. If moving to this neighbor leads to a valid SAT, then we accept this move with probability $P^+_{\rm BS} = \min\{1, (2d-1)z\}$. Otherwise, reject this move and return to Step 1.
    \item If the ``Delete" action is selected and $\scrN>0$, then delete the last edge of the trail with probability $P^{-}_{\rm BS} = \min\{1, \frac{1}{(2d-1)z}\}$. Otherwise, return to Step 1.
\end{enumerate}

We next introduce the irreversible algorithm, in which the balance condition is satisfied but the detailed balance condition is violated. In this algorithm, the state space of the SATs is doubled by introducing an auxiliary variable with two values, $+$ and $-$, to each trail. A typical state in the enlarged state space can be written as $(\omega, +)$ or $(\omega, -)$, where $\omega$ is a trial. The simulation procedure for the irreversible algorithm is described as follows.
\begin{enumerate}
    \item In the increasing mode $+$ (i.e., at a state $(\omega, +)$), uniformly at random choose one of the $2d-1$ neighboring sites of the head. If moving to this neighbor leads to a valid SAT, then accept this action with probability $P^+ = \min\{1, (2d-1)z\}$. Otherwise, switch to the decreasing mode $(-)$, i.e., updating $(\omega, +)$ to $(\omega, -)$.
    \item In the decreasing mode $(-)$ (i.e., at a state $(\omega, -)$), delete the last edge of the trail with probability $P^{-} = \min\{1, \frac{1}{(2d-1)z}\}$ if $\scrN>0$. Otherwise, switch to the increasing mode $(+)$, i.e., updating $(\omega, -)$ to $(\omega, +)$.
\end{enumerate}


For SATs in square and simple cubic lattices, near critical points the value of $(2d-1)z$ is slightly larger than $1$. It follows that $P^+=1$ and $P^- \lesssim 1$ for the range of $z$ we are interested in. Therefore, in simulations, a trail in the increasing mode keeps growing until it intersects with itself. After that, it switches to the decreasing mode $(-)$ and starts to delete edges from the head of the trail with probability $P^-$ which is close to 1. If the deletion is rejected, it switches back to the increasing mode and continues to grow. Since $P^{+}=1$ and $P^{-}\lesssim 1$, the irreversible algorithm can efficiently update trails.

\section{OBSERVABLES AND Finite-size scaling}
\label{Sec:Observables}
In simulations, we sample the length of self-avoiding trails $\scrN$, and define an indicator $\scrD_0$ for zero-length SATs, i.e., $\scrD_0 = \mathbf {1}({\scrN = 0})$. Denote a trail on the lattice with length $\scrN$ by $\omega = (\omega_0, \omega_1, \cdots, \omega_\scrN)$ with each $\omega_i \in \ZZ^d_L$. We also measure the end-to-end distance $\scrR = \|\omega_\scrN - \omega_0\|$. In addition, we sample an \emph{unwrapped} end-to-end distance $\scrR_u$, which is the end-to-end distance if the trail is embedded from the torus to the infinite lattice. More explicitly, for a given SAT, the unwrapped end-to-end distance $\scrR_u$ can be calculated by traversing the trail as follows. Starting from one endpoint of the trail (say $\omega_0$) with a vector $U\in \ZZ^d$ set to 0, and then traverse along the trail to the end. In each step, the vector is updated as $U = U + e_i$ ($-e_i$) if the traverse is along (against) the $i$-th direction. Here $e_i$ is the unit vector of the $i$-th direction. After the traverse is finished, one has $\scrR_u = \|U\|$.

We then calculate the ensemble average (denoted as $\langle \cdot \rangle$) of the following observables.
\begin{enumerate}
    \item The mean trail length $N = \langle \scrN \rangle$, and its variance $C=\frac{1}{V}(\langle\mathcal{N}^2\rangle-{\langle\mathcal{N}\rangle}^2)$ where $V=L^d$ is the volume.
    \item The susceptibility $\chi=1/{\langle \mathcal{D}_0 \rangle}$. This equality holds since $\langle \mathcal{D}_0\rangle= \frac{\sum_{\omega:|\omega|=0 }z^{|\omega|}}{\sum_{\omega}z^{|\omega|}}=1/\chi$.
    \item The mean unwrapped end-to-end distance $\xi_u = \langle\scrR_u\rangle$.
    \item A dimensionless Binder ratio $Q=\frac{\langle \mathcal{N}^2\rangle}{{\langle\mathcal{N}\rangle}^2}$.
\end{enumerate}

The finite-size scaling of the above quantities near the critical point are expected as follows.
\begin{equation}
\label{Eq:FSS-formula}
    \begin{aligned}
        N(z,L)&=L^{\frac{1}{\nu}}N_s[(z-z_c)L^{y_t}],\\
        \chi(z,L)&=L^{\frac{\gamma}{\nu}}\chi_s[(z-z_c)L^{y_t}],\\
        \xi_u(z,L)&=L\xi_{us}[(z-z_c)L^{y_t}],\\
        C(z,L)&=L^{\frac{\alpha}{\nu}}C_s[(z-z_c)L^{y_t}],\\
        Q(z,L)&=Q_s[(z-z_c)L^{y_t}],
    \end{aligned}
\end{equation}
where $N_s(\cdot),\chi_s(\cdot),\xi_{us}(\cdot),C_s(\cdot),Q_s(\cdot)$ are scaling functions. It is clear from Eq.~\eqref{Eq:FSS-formula} that, at the critical point, the scaling functions are constant and therefore the quantities $N$, $\chi$, $\xi_u$ and $C$ scale as a power-law with respect to $L$ while $Q$ approaches to a constant. 


\section{Results}
\label{sec:re}
In this section, we perform the least-squares fits of the sampled quantities to the fitting ansatz from the Taylor expansion of the finite-size scaling in Eq.~\eqref{Eq:FSS-formula}. To account for the finite-size corrections, we introduce a minimum system size $L_{\rm min}$ and include only data with $L\geq L_{\rm min}$ in the fits. We then gradually increase $L_{\rm min}$ to check the effect of small system data on the estimates of various parameters in the fitting ansatz. Our preferred fitting is the one with ${\rm chi}^2/{\rm DF} \lesssim 1$ where ${\rm chi}^2$ is the residual and {\rm DF} is the degree of freedom, and the subsequent increase of $L_{\rm min}$ does not cause a significant drop in ${\rm chi}^2$ compared to the decrease of ${\rm DF}$.

\subsection{Estimate of $z_c$ on the simple cubic lattice}
In this section, we analyze the data of $\xi_u$ and $Q$ to estimate the critical point $z_c$ on the simple cubic lattice. For the SAW model, these quantities have been shown to exhibit weaker finite-size corrections and are expected to yield high-precision estimates of $z_c$. Since both $\xiu/L$ and $Q$ are dimensionless, we fit the data of $\xi_u/L$ and $Q$ to the following ansatz,
\begin{equation}
\label{Eq:ansatz_near_zc}
    \begin{aligned}
       O & = q_0 + \sum_{k=1}^{m} q_k(z-z_c)^kL^{k y_t} + b_1L^{y_1},
    \end{aligned}
\end{equation}
where $O$ represents $\xi_u/L$ and $Q$, $m$ is the highest order retained in the Taylor expansion, and the term $b_1L^{y_1}$ accounts for the finite-size corrections with $y_1 < 0$. Other correction terms commonly appearing in the finite-size scaling, such as lower order finite-size corrections $b_2L^{y_2}$ with $y_2 < y_1$ and mixed correction term $c_1(z-z_c)L^{y_t+y_1}$ are not clearly detected in our data for these quantities. Thus, we omit these terms from the fitting ansatz.

\begin{table*}
\caption{\label{tab:table3}Fitting results for the unwrapped end-to-end distance $\xi_u$ and the Binder ratio $Q$ using the ansatz Eq.~\eqref{Eq:ansatz_near_zc} with $y_1$ fixed at $-1$, on the simple-cubic lattice.}

\begin{ruledtabular}
\begin{tabular}{ccccccccccc}
 \textit{O}&$L_{\rm min}$&$y_t$&$z_c$&$q_0$&$q_1$&$q_2$&$q_3$&$q_4$&$b_1$ &${\rm chi}^2/{\rm DF}$\\ \hline
$\xi_u$
           &32&1.684(4)&0.206\,377\,11(7) &0.485\,4(4)&-0.89(2)&0.59(8)&1.9(7)&23(8)&0.28(1)&47.6/37\\
           &48&1.688(5)&0.206\,377\,00(9) &0.484\,7(5)&-0.87(2)&0.54(7)&2.0(6)&23(8)&0.32(3)&31.5/30\\
           &64&1.699(7)&0.206\,376\,89(11)&0.483\,7(9)&-0.82(3)&0.53(8)&1.6(6)&15(7)&0.38(6)&17.1/23\\[3pt]
    $Q$    
           &32&1.691(5)&0.206\,376\,93(7) &1.408\,3(5)&1.21(2)&-0.02(9)&-3(1)&0(9)&0.22(2)&51.5/38\\
           &48&1.698(5)&0.206\,376\,94(10)&1.408\,2(8)&1.17(3)&0.01(9)&-3.2(9)&-1(9)&0.22(4)&28.8/31\\
           &64&1.707(7)&0.206\,376\,83(13)&1.410(2)&1.11(4)&-0.05(9)&-2.4(9)&4(8)&0.10(9)&17.3/24\\[3pt]
\end{tabular}
\end{ruledtabular}
\label{Tab:3D-fit-zc}
\end{table*}

\begin{figure}[htb]
    \centering
    \includegraphics[width=1.0\linewidth]{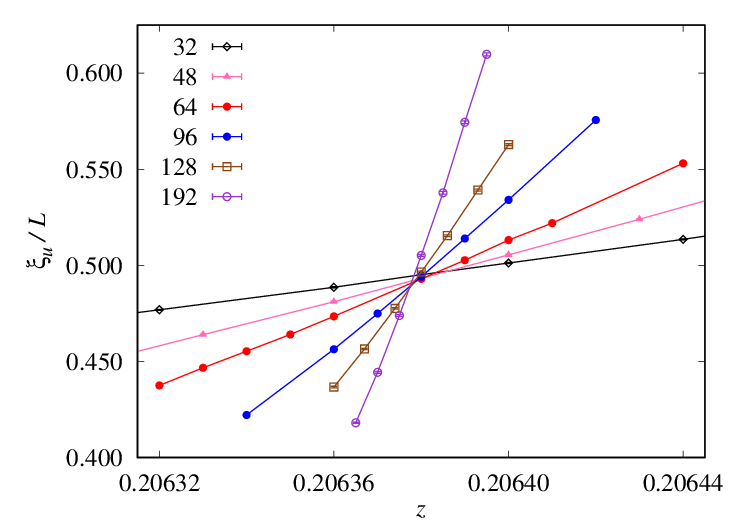}
    \caption{Plot of the unwrapped end-to-end distance $\xi_u$ rescaled by $L$, on the simple-cubic lattice.}
    \label{Fig:xiu-3d}
\end{figure}

We first discuss the data of $\xi_u$ in the simple cubic lattice. In Fig.~\ref{Fig:xiu-3d}, we plot the data $\xiu/L$ versus the fugacity $z$. It can be clearly seen that the data for various system sizes intersect around $z = 0.206\,38$. To obtain a precise estimate of the insection point, which is $z_c$, we fit the data of $\xiu$ to the ansatz in Eq.~\eqref{Eq:ansatz_near_zc}, where $O$ denotes $\xiu/L$. We first try to fit the data with $m=4$ and without the correction term, i.e., setting $b_1 = 0$. However, it shows that the ratio ${\rm chi}^2/{\rm DF}$ is still large even when we set $L_{\rm min} = 96$. This indicates the existence of strong finite-size corrections. We then include the correction term $b_1L^{y_1}$ in the fitting ansatz. However, leaving both $b_1$ and $y_1$ free in the ansatz leads to unstable results. We then try to fix $y_1 = - 1$ and leave $b_1$ free. The fitting result is stable when $\Lmin = 48$, which produces $z_c = 0.206\, 377\,00(9)$ and $y_t = 1.688(5)$. Increasing $\Lmin$ to $64$ produces consistent fitting results. The fitting details are shown in Table ~\ref{Tab:3D-fit-zc}. We also try to fit by fixing $y_1$ to $-2$ or $-1/2$, but both fits produce consistent estimates of $z_c$. Consistent estimates are also obtained when including two correction terms $b_1L^{-1} + b_2L^{-2}$ in the fitting ansatz. In three dimensions, although the exact value of $y_t$ is not known, a precise estimate for $\nu$ is available. It was estimated that $\nu=0.587\,597\,00(40)$ and thus $y_t = 1/\nu$ is centered around $1.701\,846\,67$. So we also try to fit by fixing $y_t$ to this value, but the fitting result is almost the same as leaving $y_t$ free. 

\begin{figure}[htb]
    \centering
    \includegraphics[width=1.0\linewidth]{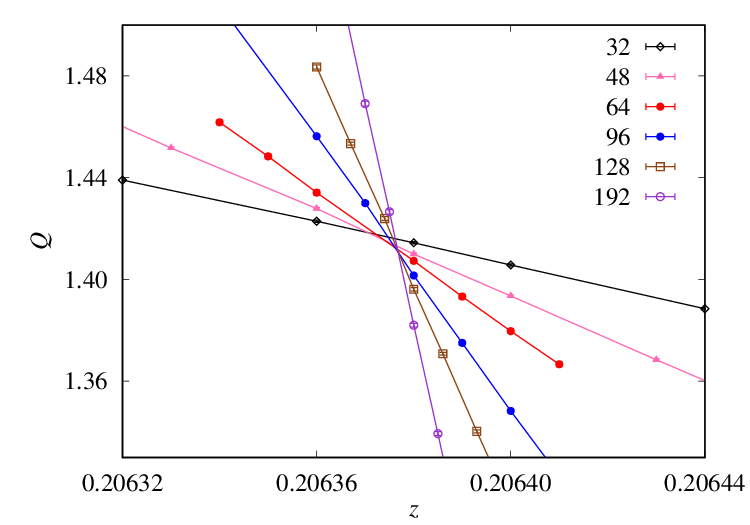}
    \caption{Plot of the Binder ratio $Q$ versus $z$ on the simple-cubic lattice.}
    \label{Fig:Q-3d}
\end{figure}

We then try to estimate $z_c$ from the data of $Q$. In Fig.~\ref{Fig:Q-3d}, we plot the data $Q$ versus the fugacity $z$ for various system sizes. Again, the data exhibit a clear intersection point around $z=0.206\,38$. We then fit the data of $Q$ to the ansatz in Eq.~\eqref{Eq:ansatz_near_zc}, where $O$ denotes $Q$. We first perform fits with $m=4$ and without any correction terms, which produces stable fitting results when $\Lmin = 64$, yielding $z_c = 0.206\,376\,71(7)$ and $y_t = 1.709(8)$. We then include the correction term $b_1L^{y_1}$ in the fitting ansatz. Again, leaving $y_1$ as a free parameter in the ansatz leads to unstable fits. After fixing $y_1$ to $-1$ and leaving $b_1$ free, stable results are obtained when $\Lmin = 48$ which produces $z_c = 0.206\,376\,94(10)$ and $y_t =  1.698(5)$. Both are consistent with previous estimates. Consistent estimates are obtained when $y_1$ is fixed to $-1/2$ and $-2$. We also try to fit by fixing $y_t$ to $1.701\,846\,67$, and it shows a negligible effect on the estimates of $z_c$ and $y_t$. To take into account the systematic error, we compare various estimates of $z_c$ from $\xiu$ and $Q$ and finally estimate $z_c = 0.206\,376\,9(2)$. 

\begin{figure}[htb]
    \centering
    \includegraphics[width=1.0\linewidth]{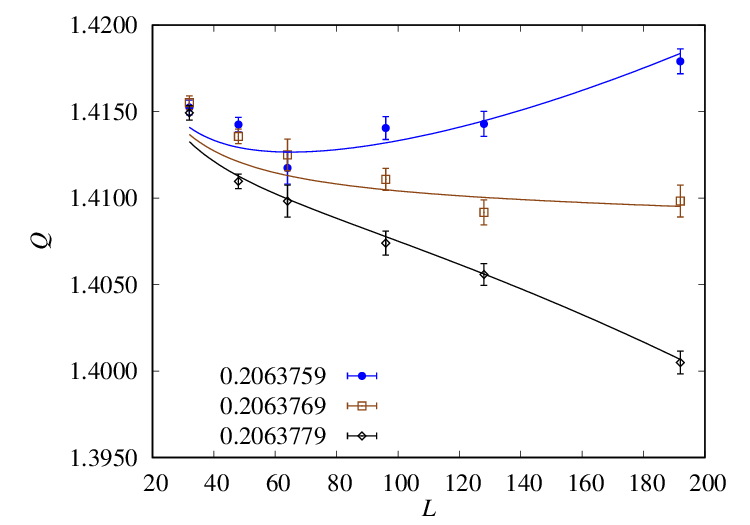}
    \caption{Plot of $Q$ versus $L$ at our estimated $z_c$ and at five standard deviations above and below our estimated $z_c$. The curves correspond to our preferred fits of the data by the ansatz Eq.~\eqref{Eq:ansatz_near_zc} on the simple cubic lattice.}
    \label{Fig:Qni-3d}
\end{figure}

To demonstrate the validity of our estimate, we plot $Q$ versus $L$ at our estimated $z_c$ and at $z_c$ plus or minus 5 standard deviations. As shown in Fig.~\ref{Fig:Qni-3d}, as $L$ increases, at $z_c$, the data of $Q$ tends to a constant, while overestimated or underestimated values of $z_c$ cause $Q$ to bend upward or downward, respectively. This clearly demonstrates the reliability of our estimate of $z_c$. Compared with the previous best estimate 0.206\,48(13), our estimate improves the precision by a factor of 500.

\subsection{Estimate of $z_c$ on the square lattice}

\begin{figure}[htb]
    \centering
    \includegraphics[width=1.0\linewidth]{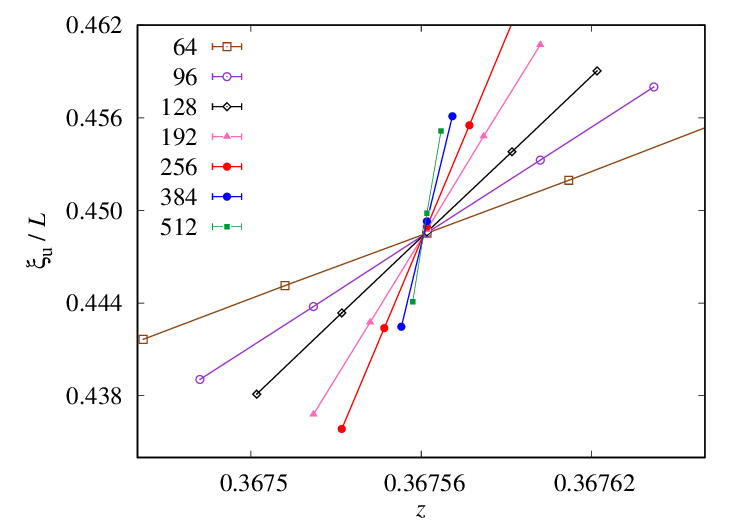}
    \caption{Plot of the unwrapped end-to-end distance $\xi_u$ rescaled by $L$ on the square lattice.}
    \label{Fig:U-2d}
\end{figure}

In this section, we study the critical point $z_c$ on the square lattice by analyzing the data of $\xiu$ and $Q$. We first discuss the data of $\xiu$ on the square lattice. In Fig.~\ref{Fig:U-2d}, we  plot the data $\xiu/L$ versus the fugacity $z$ for various system sizes, and a clear intersection can be seen around $z = 0.367\,56$. We next fit the data of $\xiu$ to the ansatz in Eq.~\eqref{Eq:ansatz_near_zc}. The fitting shows that, even with $m=1$ and without correction terms, the data can be well fitted when $\Lmin = 96$, producing $z_c = 0.367\,560\,80(4)$ and $y_t = 1.335(3)$. The exponent is consistent with the exact SAW exponent $4/3$. Setting $m=2$ leads to almost the same results, with $q_2$ being consistent with zero. We then include correction terms in the ansatz. Again, including the term $b_1 L^{y_1}$ with $y_1$ free leads to unstable fits. We then fix $y_1  = - 1$, and the results are shown in Table ~\ref{Table:fit-2d}. The fits are stable when $\Lmin = 96$, giving $z_c = 0.367\,560\,89(7)$ and $y_t = 1.335(3)$. The value of $b_1$ is quite small, implying that finite-size corrections are weak, which is consistent with the clear intersection in Fig.~\ref{Fig:U-2d}. We also try to fit by fixing $y_t = 4/3$, which is the exact SAW exponent in two dimensions. Again, the result shows that fixing $y_t$ or not has almost no effect on the estimates of $z_c$ and $y_t$.

Finally, we study the data of $Q$ on the square lattice. We first try to fit the data with $m=2$ and without correction terms. The fits show that the residual is still large when $\Lmin = 128$. Setting $m = 3$ in the fitting ansatz cannot reduce the residual significantly, and the coefficient $q_3$ is consistent with zero. This implies the existence of strong finite-size corrections. We then fit with $m=2$ and one correction term $b_1L^{y_1}$. Leaving $y_1$ free still cannot produce stable fits. Fixing $y_1 = -1$ leads to stable fits when $\Lmin = 96$, yielding $z_c = 0.367\,561\,08(6)$ and $y_t = 1.339(3)$, which are consistent with the estimates from $\xiu$. Adding another correction term $b_2 L^{-2}$ to the ansatz produces consistent results, but $b_2$ is consistent with zero. Again, consistent fitting results are obtained when we fix the exponent $y_t = 4/3$ in the fitting ansatz.

\begin{figure}[htb]
    \centering
    \includegraphics[width=1.0\linewidth]{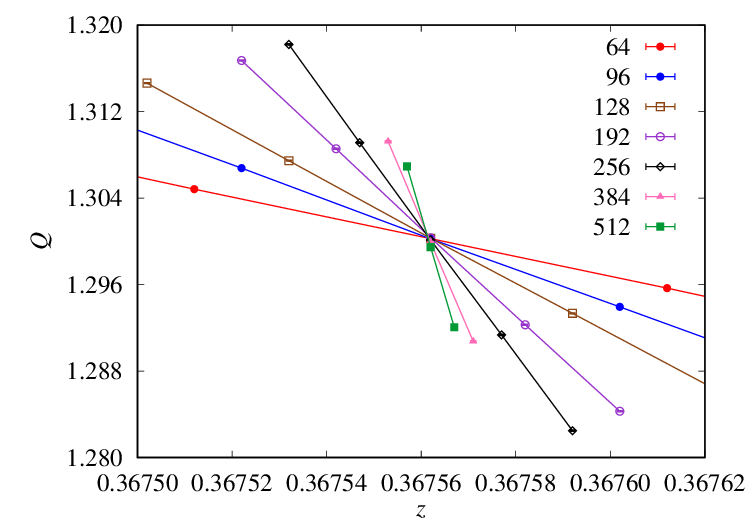}
    \caption{Plot of the Binder ratio $Q$ versus $z$ on the square lattice.}
    \label{Fig:Q-2d}
\end{figure}

\begin{figure}[htb]
    \centering
    \includegraphics[width=1.0\linewidth]{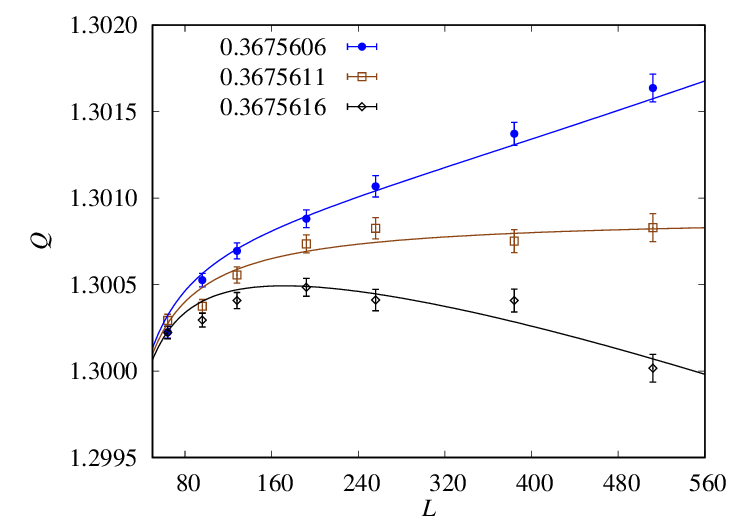}
    \caption{Plot of the data of $Q$ versus $L$ at our estimated $z_c$, and at five standard deviations above and below $z_c$. The curves correspond to our preferred fit of $Q$ by the ansatz Eq.~\eqref{Eq:ansatz_near_zc} on the square lattice.}
    \label{Fig:Qni-2d}
\end{figure}

\begin{table*}
\caption{\label{tab:table3}Fitting results for the unwrapped end-to-end distance $\xi_u$ and the Binder ratio $Q$ using the ansatz Eq.~\eqref{Eq:ansatz_near_zc} with $y_1$ fixed at $-1$, on the square lattice.}
\begin{ruledtabular}
\begin{tabular}{ccccccccc}
 \textit{O}&$\Lmin$&$y_t$&$\zc$&$q_0$&$q_1$&$q_2$&$b_1$&${\rm chi}^2/{\rm DF}$\\ \hline
$\xiu$
  &64&1.334(2)&0.367\,560\,73(5) &0.448\,30(4)&-0.268(3)&0.01(1)&0.009(3)&36.2/25\\
  &96&1.335(3)&0.367\,560\,89(7) &0.448\,48(7)&-0.268(4)&0.004(12)&-0.009(6)&17.7/20\\
  &128&1.332(4)&0.367\,560\,99(9)&0.448\,6(1) &-0.272(6)&0.004(14)&-0.03(1)&12.7/15\\[3pt]

\textit{Q}
 &64&1.340(2)&0.367\,561\,19(5)&1.300\,79(5)&0.353(3)&0.08(2)&-0.032(4)&41.4/25\\
 &96&1.339(3)&0.367\,561\,08(6)&1.300\,96(8)&0.355(5)&0.08(2)&-0.050(7)&22.2/20\\
 &128&1.337(4)&0.367\,561\,10(8)&1.300\,9(2)&0.359(7)&0.08(2)&-0.05(2)&15.9/15\\[3pt]
 
\end{tabular}
\end{ruledtabular}
\label{Table:fit-2d}
\end{table*}
 Finally, by comparing various fitting results, we estimate $z_c = 0.367\,561\,1(1)$ for SATs on the square lattice,  which improves the previous best estimate by a factor of 70. To validate the reliability of our estimate of $z_c$, we plot $Q$ versus $L$ at our estimated $z_c$, and at $z_c$ plus or minus 5 standard deviations, as shown in Fig.~\ref{Fig:Qni-2d}. As expected, the values of $Q$ at $z_c$ tend to a horizontal line, while the data above and below $z_c$ clearly bend upward and downward.

\subsection{Scaling behavior at $z_c$ on the simple cubic lattice}

This section focuses on the scaling behaviors of the mean trail length $N$, its variance $C$ and the susceptibility $\chi$, at the critical point. From Eq.~\eqref{Eq:FSS-formula}, the expected scaling behaviors of these quantities at $z_c$ are $N\sim L^{1/\nu}$, $C \sim L^{\alpha/\nu}$ and $\chi \sim L^{\gamma/\nu}$, respectively. To estimate these critical exponents, we fit the data of $N$, $C$, and $\chi$ to the following ansatz,
\begin{equation}
    \scrO = L^{y_\scrO}(a_0+a_1L^{y_1}+a_2L^{y_2})+c_0.
    \label{Eq:ansatz_zc}
\end{equation}
For $N,\chi$ and $C$, the corresponding exponents $y_\scrO$ are $1/\nu$, $\gamma/\nu$ and $\alpha/\nu$, respectively.

We first fit the data of $N$ to Eq.~\eqref{Eq:ansatz_zc} without correction terms, i.e., setting $a_1 = a_2 = 0$. The fits are stable when $\Lmin = 32$. From the fitting, we estimate $y_t = 1/\nu = 1.699(4)$ or $\nu = 0.588\,6(14)$. The fitting details are shown in Table ~\ref{Tab:fit-zc}. Including one correction term $b_1L^{-1}$ in the ansatz leads to similar results but the error bar of $y_t$ is  larger. Our estimate of $\nu$ is in excellent agreement with the SAW value $\nu=0.587\,597\,00(40)$~\cite{clisby2007self}. Similarly, the data for $\chi$ are fitted to the ansatz to estimate the exponent $\gamma/\nu$. Without correction terms, the fits are stable when $\Lmin = 32$ and we have $\gamma /\nu = 1.961(5)$. Using the estimate of $\nu$, we obtain $\gamma = 1.154(6)$, which is consistent with the accurate SAW value $\gamma=1.156\,953\,00(95)$ in Ref.~\cite{clisby2007self}. An analogous fit is performed for the variance of the mean trail length $C$ to estimate the exponent $\alpha/\nu$. Again, only including $a_0$ and $c_0$ terms in the fitting ansatz leads to stable fits when $\Lmin = 32$, and we estimate $\alpha/\nu = 0.43(2)$, and thus $\alpha = 0.253(13)$. Using the precise estimate of $\nu$ from SAW~\cite{clisby2007self} and the scaling relation $\alpha = 2 - \nu d$, one can obtain that $\alpha = 0.237\,209\,0(12)$. Thus, within two standard deviations, our estimate of $\alpha$ for SAT is also consistent with SAW.

\begin{figure}[htb]
    \centering
    \includegraphics[width=1.0\linewidth]{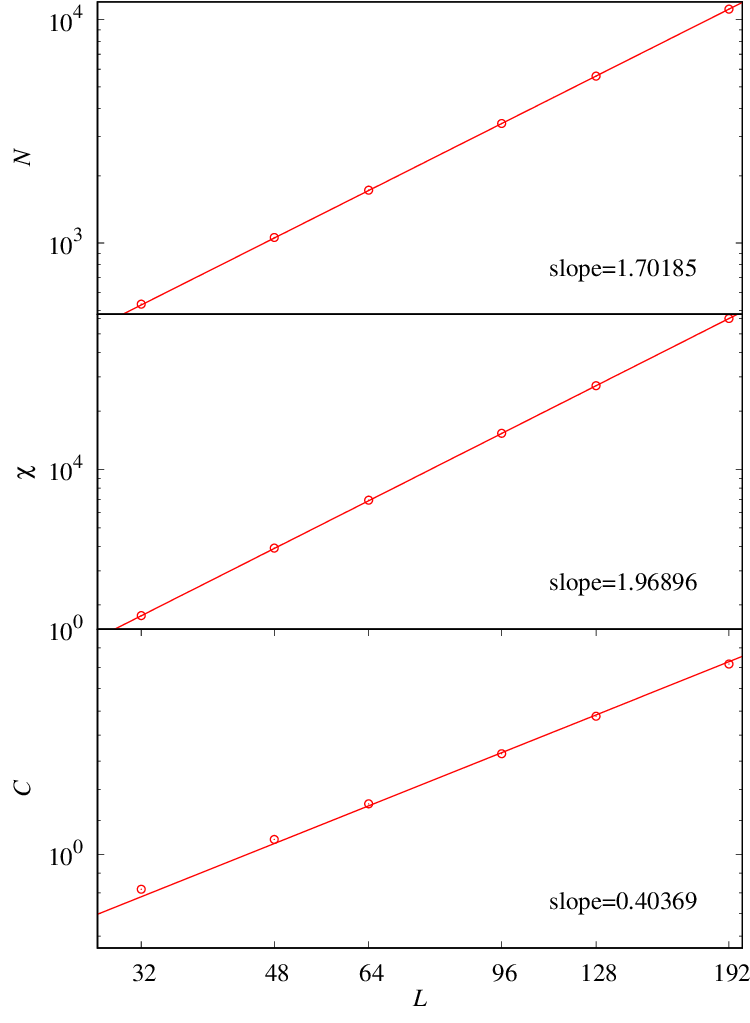}
    \caption{Plots of $N, \chi, C$ versus $L$ at $z=0.206\,376\,9$ with different system sizes. The slopes of the straight lines in the top, middle and bottom plots are respectively the value of $1/\nu$, $\gamma/\nu$ and $\alpha/\nu$, these critical exponents are the most precise results from the SAW. }
    \label{Fig:exp-3d}
\end{figure}

To clearly show that the scaling behaviors of $N$, $\chi$, and $C$ for SAT are consistent with SAW, we plot in Fig.~\ref{Fig:exp-3d} the data of these quantities using log-log scale. The slopes of the straight lines in the top, middle and bottom plots are respectively the value of $1/\nu$, $\gamma/\nu$ and $\alpha/\nu$ from SAW. The excellent collapse of data points onto these lines suggests that SAT and SAW share the same set of exponents.


\begin{figure}[!htb]
    \centering
    \includegraphics[width=1.0\linewidth]{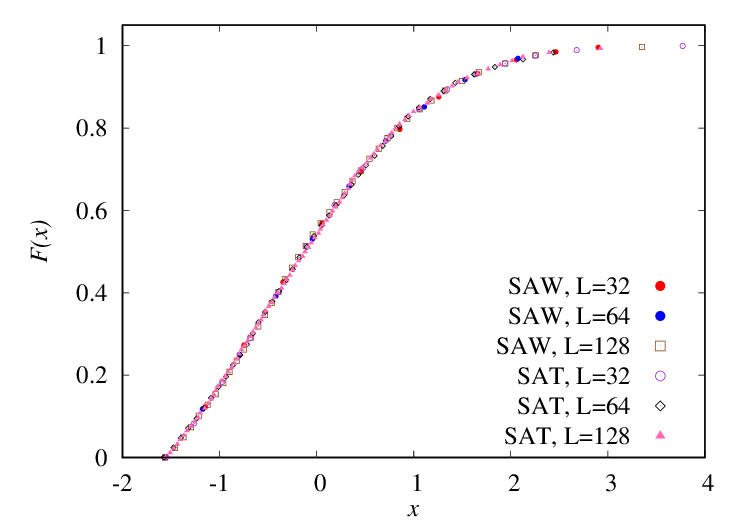}
    \caption{Plot of critical probability distributions for the trial length of SAT and walk length of SAW on simple-cubic lattice, with system sizes $L=32,64,128$. }
    \label{Fig:walk-length-distribution}
\end{figure}

\begin{table}[htb]

\caption{\label{tab:table2}
    The fitting results of the susceptibility $\chi$, the mean trail length $N$ and its variance $C$ at the critical point $z_c=0.206\,376\,9$ on the simple cubic lattice.}
\begin{ruledtabular}
\begin{tabular}{cccccc}
O&$L_{\rm min}$&$y_t$&$a_0$&$c_0$&$\rm{chi}^2/\rm{DF}$\\ \hline
$N$ &32 &1.699(2) &1.47(2) &3(2) &2.1/3\\
    &48 &1.703(4) &1.44(3) &9(5) &0.3/2\\
    \hline 
 $\chi$&32&1.961(5)&1.99(5)&-20(14)&0.3/3\\
      &48&1.962(10)&1.98(10)&-14(48)&0.3/2\\
      \hline
$C$ &32&0.43(2)&0.72(6)&0.5(2)&1.5/3\\
    &48&0.44(3)&0.6(1)&0.7(3)&0.7/2\\
\end{tabular}
\end{ruledtabular}
\label{Tab:fit-zc}
\end{table}
We next study the length distribution of SAT and SAW on simple-cubic lattice, at the critical point. Even the quantities $N$, $C$, and $\chi$ exhibit the same scaling behaviour for SAT and SAW, it does not imply that these two models have the same length distribution. We thus sample the trail length of the SAT at $z=0.206\,376\,9$ and the walk length of the SAW at $z=0.213\,491$, both with system sizes $L=32,64,128$. For both models, we study the standardized probability distribution function, defined as follows,
\begin{equation}
    \label{Eq:F(x)}
    F(x):=\PP\left(\frac{ \scrN - \EE\scrN}{\sqrt{\var(\scrN)}}\leq x\right).
\end{equation}
In Fig.~\ref{Fig:walk-length-distribution}, we plot the data of $F(x)$ for SAT and SAW. It clearly suggests that, at the critical point, the walk and trail length distributions for SAT and SAW converge to the same scaling limit.

\subsection{Efficiency of the algorithms}
\label{sec:eff-algo}
The efficiency of the BS and irreversible algorithms on the self-avoiding walk model in dimensions from 2 to 5 was systematically studied in Ref.~\cite{hu2017irreversible}. On the simple cubic lattice, the autocorrelation time of the walk length for the BS algorithm is approximately 13 times larger than that of the irreversible algorithm. Namely, for the same simulation time, the standard deviation of the walk length from the irreversible algorithm is approximately $\frac{1}{\sqrt{13}}$ of that in the BS algorithm, which clearly demonstrates the efficiency of the irreversible algorithm.

In this work, we also study and compare the efficiency of the two algorithms on the SAT model, and we focus on the simple-cubic lattice. Compared with SAW, the BS algorithm could be less effective to simulate SAT. This is because the length of a self-avoinding trail could be much longer than a self-avoiding walk, and it takes a longer time for the BS algorithm to significantly update a long trail. But it is not a problem for the irreversible algorithm, since the trail keeps growing in the increasing mode until it revisits a visited edge and in the decreasing mode it deletes edges from the head with a high probability. So we expect the advantage of the irreversible algorithm to be more obvious for SAT.

To compare the efficiency, we study the standard deviations of the mean trail length obtained from the two algorithms, under similar CPU time. The data are shown in Table~\ref{tab:efficiency-compare}. It can be seen that the standard deviation from the irreversible algorithm is $\approx \frac{1}{6}$ of that from the BS algorithm. In terms of the autocorrelation time, this implies that the former is $\approx \frac{1}{36}$ of the autocorrelation time of the BS algorithm. This demonstrates that, for the SAT model, the advantage of the irreversible algorithm over the BS algorithm is more obvious than in the SAW model.


\begin{table}[htb]
\caption{\label{tab:efficiency-compare}
Comparison of the standard deviations of the mean trial length on the simple-cubic lattice and at the critical point, respectively obtained using the BS algorithm and the irreversible algorithm.}
\begin{ruledtabular}
\begin{tabular}{ccccc}
Algorithm&$L$&$N$&$\delta \scrN$& CPU time/min\\ \hline
IR algorithm          &32&532.57&1.97&0.48\\
                      &48&1058.16&3.37&2.35\\
                      &64&1732.35&5.32&7.24\\
                      &96&3438.14&9.48&37.03\\
                      &128&5613.81&14.02&116.38\\
BS algorithm  &32&547.21&11.79&0.45\\
              &48&1101.57&22.36&2.51\\
              &64&1715.44&35.86&7.54\\
              &96&3425.27&57.35&38.95\\
              &128&5515.47&82.25&116.48\\
\end{tabular}
\end{ruledtabular}
\end{table}

\section{Conclusion}
\label{Sec:Conclusion}

In this work, we systematically studied the critical properties of self-avoiding trails on the two-dimensional square lattice and the three-dimensional simple cubic lattice with periodic boundary conditions. By performing finite-size analysis of the unwrapped end-to-end distance $\xi_u$ and the Binder ratio $Q$, we accurately determined the critical points of SATs. On the simple cubic lattice, we obtained $z_c=0.206\,376\,9(2)$, which improves the precision of the best estimate of the existing by a factor of 500. On the square lattice, we estimate $z_c=0.367\,561\,1(1)$, which improves the precision of the best existing result by a factor of 70.  

At the estimated $z_c$ of the simple cubic lattice, we further analyzed the scaling behaviors of various observables, including the mean trail length and its variance, and the susceptibility. We estimated the critical exponents $\nu=0.589\,3(7)$, $\gamma=1.158(2)$ and $\alpha = 0.253(13)$, which are consistent with those of the SAW model in three dimensions. Moreover, at $z_c$, our data suggest that the probability distributions of the walk/trail length for these two models converge to the same scaling limit.

In general, simulating SATs on high-dimensional lattices could be challenging, since the trail can grow significantly long. This work demonstrates that the irreversible algorithm can potentially be used to efficiently simulate the SAT model in high dimensions. In recent years, the finite-size scaling of the SAW model above the upper critical dimension ($d > 4$), has attracted much attention since the scaling behaviors of many quantities are boundary-dependent; a brief review can be found in the Introduction of Ref.~\cite{fang2021logarithmic}. It would be interesting to employ the irreversible algorithm to systematically study the finite-size scaling of the SAT model in dimensions above 4. 


\bibliographystyle{apsrev4-1}
\bibliography{apssamp}

\end{document}